\documentclass{iopart}
\usepackage{iopams}

\expandafter\let\csname equation*\endcsname\relax

\expandafter\let\csname endequation*\endcsname\relax

\usepackage{amsmath}
\usepackage{epsf,amssymb,fleqn}
\catcode`\@=11
\usepackage{epsfig}
\usepackage{epstopdf}
\usepackage{float}
\usepackage{color,colordvi}
 \usepackage{cite}

\def\II{\hbox{{1}\kern-.25em\hbox{l}}}

\begin{document}
\begin{flushright}
{{{{{{{{\textsl{DESY-16--230}}}}}}}}}
\end{flushright}

\title[Spin Chains and Gustafson's Integrals ]{Spin Chains and Gustafson's Integrals }

\author{S. {\'E}.~Derkachov$^{1}$
and A. N.~Manashov$^{2,3}$ }

\address{$^1$ St.Petersburg Department of Steklov
Mathematical Institute of Russian Academy of Sciences,\\
\ \ \
Fontanka 27, 191023 St.~Petersburg, Russia.
}

\address{$^2$  Institut f\"ur Theoretische Physik, Universit\"at Hamburg,   D-22761 Hamburg, Germany}
\address{$^3$ Institute for Theoretical Physics, University of  Regensburg, D-93040 Regensburg, Germany}

\ead{derkach@pdmi.ras.ru}
\ead{alexander.manashov@desy.de}
\begin{abstract}
Gustafson's integrals are  multidimensional generalizations of the classical Mellin--Barnes  integrals. We show
that some of these integrals arise from relations between  matrix elements in  Sklyanin's representation of
Separated Variables in  spin chain models. We also present  several new integrals.
\end{abstract}

\maketitle

\setcounter{footnote}{0}

\section{Introduction}
In the papers~\cite{Gustafson,Gustafson92} R.A. Gustafson has calculated integrals representing   multidimensional
generalization of the Mellin--Barnes integrals. The integrals associated with the classical
 $\text{su}(N)$ and  $\text{sp}(N)$ Lie
algebras take the form~\cite{Gustafson}
\\
\begin{eqnarray}\label{Gustafson-I}
 \fl \left(\prod_{n=1}^{N}\int_{-i\infty}^{i\infty} \frac{dz_n}{2\pi i} \right)\,
\frac{\prod_{k=1}^{N+1} \prod_{j=1}^{N} \Gamma(\alpha_k-z_j)\Gamma(\beta_k+z_j)}{
\prod_{1\leq k< j\leq N}\Gamma(z_k-z_j)\Gamma(z_j-z_k)}=
\frac{N!\prod_{k,j=1}^{N+1} \Gamma(\alpha_k+\beta_j)}{\Gamma\big(\sum_{k=1}^{N+1}(\alpha_k+\beta_k )\big)}
\end{eqnarray}
\\
and
\\
\begin{eqnarray}\label{Gustafson-II}
\fl  \left(\prod_{n=1}^{N}\int_{-i\infty}^{i\infty} \frac{dz_n}{2\pi i} \right)\,
\frac{\prod_{k=1}^{2N+2} \prod_{j=1}^{N} \Gamma(\alpha_k \pm z_j)}{\prod_{k=1}^N\Gamma(\pm 2z_k)
\prod_{1\leq k< j\leq N}\Gamma(z_k \pm z_j)\Gamma(-z_k \pm z_j)}=
\frac{2^NN!\prod_{k<j} \Gamma(\alpha_k+\alpha_j)}{\Gamma\big(\sum_{k=1}^{2N+2}\alpha_k \big)}\,,
\end{eqnarray}
\\
where $\Gamma(\alpha\pm \beta)\equiv \Gamma(\alpha+\beta)\Gamma(\alpha-\beta)$ and the integration contours separate
the series of poles of $\Gamma$ functions, $\{\alpha_k +n_k\}$  and  $\{-\beta_k-n_k\}$, $k=1,\ldots ,N$, $n_k\in
\mathbb{Z}_+$,
 in the first integral, and $\{\alpha_k+n_k\}$ and $\{-\alpha_k-n_k\}$, $k=1,\ldots, 2N+2$ in
the second one.

Gustafson's integrals and their $q$- and $p,q$-
generalization~\cite{Gustafson,Gustafson92,Gustafson94,Spiridonov,Spiridonov06,Spiridonov07} play an important role in many
topics in physics and mathematics such as theory of multivariable orthogonal polynomials~\cite{Stokman}, Selberg type
integrals and constant term identities~\cite{Andrews-Askey-Roy,Forrester-Warnaar}, supersymmetric dualities in quantum field
theory~\cite{Spiridonov-Vartanov}.

The aim of this paper is to demonstrate that Gustafson type integrals arise in  a natural way in integrable spin chain models.
Namely,  the integrals~\eqref{Gustafson-I} and \eqref{Gustafson-II} can be related to  matrix elements of the shift operator
$T_\gamma$ (the operator of translations) in   Sklyanin's representation  of Separated Variables
(SoV)~\cite{Sklyanin:1995bm}. Moreover, we obtain a new identity which we  were not able  to derive from
integrals~(\ref{Gustafson-I}) and (\ref{Gustafson-II}). It takes the form
\\
\begin{eqnarray}\label{Gustafson-III}
\fl \left(\prod_{n=1}^{N}\int_{-i\infty}^{i\infty} \frac{dz_n}{2\pi i} \right)\, \frac{ \prod_{j=1}^{N}
\left(\prod_{k=1}^{N+1} \Gamma(\alpha_k-z_j)\right)\left( \prod_{m=1}^{N}\Gamma(z_j\pm \beta_m)\right)}{
\prod_{k<j}\Gamma(z_k\pm z_j)\Gamma(z_j-z_k)} =
\frac{ N!\prod_{j=1}^{N}\prod_{k=1}^{N+1}
\Gamma(\alpha_k\pm\beta_j)}{\prod_{1\leq j<k\leq N+1} {\Gamma(\alpha_j+\alpha_k)} }\,,
\end{eqnarray}
\\
where it is assumed that the series of poles $\{\alpha_k +n_k\}$ and $\{\pm \beta_k - n_k\}$ are separated by the integration
contours.

We  also calculate  the scalar products between the eigenfunctions of  elements of monodromy matrix and show that
evaluation of these scalar products in the SoV representation gives rise to new integral identities.

The paper is organized as follows:  Section~\ref{basic-spin-chain} contains  basic  facts about   spin chain models.
In Section~\ref{SOV} we recall the construction of the SoV representation and provide an explicit expressions for
the corresponding basis functions. Scalar products of certain  eigenfunctions  are calculated in
Sections~\ref{sect:matrix-elements} and \ref{sect:mixed}. We also show that the SoV representation for  matrix
elements  of the translation operators gives rise to the Gustafson integrals~(\ref{Gustafson-I}) and
(\ref{Gustafson-II}). In  sect.~\ref{sect:NN} we present several new integrals
 which follow from  relations between the eigenfunctions of the  monodromy matrix for closed spin chain.
The final Section~\ref{sect:summary} contains a short summary and outlook.
 Some technical details and  elements of the diagrammatic technique  employed in this paper are
given in the \ref{app:Diagram}.

\section{Spin chain models}\label{basic-spin-chain}
One dimensional quantum mechanical lattice models with  dynamical variables being generators of some Lie algebra are
usually called spin chain magnets. We consider a model with the $\text{SL}(2,\mathbb{R})$ symmetry. The dynamical
variables are the generators of this group
\begin{eqnarray}\label{SL2R-generators}
S_+^{(k)}=z_k^2\partial_{z_k}+2 s_k z_k,\hskip 10mm &  S_0^{(k)}=z_k\partial_{z_k}+ s_k, \hskip 10mm & S_-^{(k)}=-\partial_{z_k}\,,
\end{eqnarray}
where the index $k$  enumerates  the lattice sites, $k=1,\ldots, N$ and  the spin parameter  $s_k$  specifies the
representation of  $\text{SL}(2,\mathbb{R})$ group in the $k$-th  site. Henceforth we will consider  homogeneous
spin chains,
$s_1=s_2=\ldots=s_N=s$. The generators~(\ref{SL2R-generators}) act on the irreducible discrete series representation
of the $\text{SL}(2,\mathbb{R})$ group, $D_s^+$, where the spin $s$ is  a positive integer or half-integer.
This representation is realized on the  space of functions holomorphic in the upper complex  half-plane~\cite{Gelfand}.
The Hilbert space of the model is given by the direct product of vector spaces of the representation $D_s^+$ at each
site,
$\mathcal{H}_N= \prod_{k=1}^N \otimes V_s$. Thus, the span is the space of functions of $N$ complex
variables holomorphic in each variable in the upper half-plane and equipped with the invariant scalar
product~\cite{Gelfand}, which takes  the form
\begin{align}\label{scalar-product}
\big( f_1,f_2 \big) = \prod_{k=1}^N \int \mathcal{D} z_k \Big(f_1(z_1,\ldots,z_N)\Big)^\dagger f_2(z_1,\ldots,z_N)\,.
\end{align}
Here the integration goes over the upper half-plane 
$y\geq 0$, ($z=x+iy$) and the integration measure is defined as
\begin{align}\label{measure}
\mathcal{D} z =\frac{2s-1}\pi (2y)^{2s-2}  dx dy\,.
\end{align}
The scalar product~(\ref{scalar-product}) is invariant under the $\text{SL}(2,\mathbb{R})$ transformations
\begin{align}\label{Tg}
f(z_1,\ldots,z_N)\mapsto [T({g}) f](z_1,\ldots,z_N)=\left(\prod_{k=1}^N \frac1{(cz_k+d)^{2s}}\right) f(z'_1,\ldots,z'_N)\,,
\end{align}
where $g^{-1}=\begin{pmatrix}
a & b\\
c & d
\end{pmatrix}\in \text{SL}(2,\mathbb{R})$ and $z'_k=(a z_k+b)/(c z_k+d)$. The generators~(\ref{SL2R-generators}) are
anti-hermitian  with respect to  this scalar product.

The Quantum Inverse Scattering Method (QISM)~\cite{Takhtajan:1979iv,Faddeev:1979gh,Kulish:1981bi,Kulish:1981gi}
allows one  to define a physically meaningful Hamiltonian as a function of the dynamical variables,
$S_k^{(\alpha)}$,
$k=1,\ldots,N$  and  provides effective tools for solving the corresponding spectral problem. The pivotal object for
the QISM machinery is the monodromy matrix, see e.g.~\cite{Sklyanin:1991ss,Faddeev:1996iy}. It is
given by a product of Lax operators~\cite{Faddeev:1979gh}
\begin{align}
L_k(u)=u+i\begin{pmatrix}
S_0^{(k)} & S_-^{(k)}\\
S_+^{(k)} & -S_0^{(k)}
\end{pmatrix}\,.
\end{align}
For the {\it closed} and {\it open} spin chains the monodromy matrices are defined
as ~\cite{Faddeev:1979gh,Sklyanin:1988yz}:
\begin{align}
T_N^{cl}(u) &= L_1(u)L_2(u)\cdots L_N(u) =\begin{pmatrix}
A_N(u) & B_N(u)\\
C_N(u) & D_N(u)
\end{pmatrix}\,,
\\[3mm]
\mathbb{T}_N^{op}(u) &= T_N(-u) \sigma_2  T_N^{t} (u)\sigma_2 = \begin{pmatrix}
\mathbb{A}_N(u) & \mathbb{B}_N(u)\\
\mathbb{C}_N(u) & \mathbb{D}_N(u)
\end{pmatrix}\,.
\end{align}
Here $\sigma_2$ is a Pauli  matrix and   $T^t_N$  is the transposed matrix.
The matrix elements $A_N(u),\ldots, D_N(u)$ ($\mathbb{A}_N(u),\ldots, \mathbb{D}_N(u)$) are  differential operators
acting on the Hilbert space of the model. By construction they are polynomials in the spectral parameter $u$.

According to  the QISM   the entries of
$T_N^{cl}(u)$ form commuting families,
\begin{align}\label{closed-comm}
[A_N(u),A_N(v)]= [B_N(u),B_N(v)]=[C_N(u),C_N(v)]=[D_N(u),D_N(v)]=0\,.
\end{align}
For  open spin chains this property holds  for the
off-diagonal elements only,
\begin{align}\label{open-comm}
[\mathbb{B}_N(u),\mathbb{B}_N(v)]=[\mathbb{C}_N(u),\mathbb{C}_N(v)]=0.
\end{align}
It follows from Eqs.~(\ref{closed-comm}), (\ref{open-comm}) that the eigenfunctions of the operators $A_N(u),\ldots,
\mathbb{D}_N(u)$  do not depend on the spectral parameter. At the same time the corresponding eigenvalues are polynomials in
$u$. It turns out that an eigenfunction is completely determined by  its  eigenvalue.
 Therefore it is convenient to label   eigenfunctions by the roots of the corresponding eigenvalues. For
example,
if $\Psi$ is the eigenfunction of $A_N(u)$ with  eigenvalue $a_N(u)=(u-x_1)\ldots(u-x_N)$ we will denote this
eigenfunction
 by
$\Psi_{\{x_1,\ldots,x_N\}}$,
\begin{equation}
A_N(u) \Psi_{\{x_1,\ldots,x_N\}} = a_N(u) \Psi_{\{x_1,\ldots,x_N\}}=(u-x_1)\ldots (u-x_N)\Psi_{\{x_1,\ldots,x_N\}}\,.
\end{equation}
The eigenfunctions of the respective operators  provide the convenient bases
for studying  spin chain models~\cite{Sklyanin:1995bm}. All these eigenfunctions admit an explicit
representation  in the form of multi-variable integrals  which we discuss in the next section.

Closing this section we note that the operators, $B_N$ and  $C_N$, $A_N$ and $D_N$, $\mathbb{B}_N$ and
$\mathbb{C}_N$ are related to each other by an inversion~\cite{Derkachov:2014gya}, so that is is sufficient to
consider  the operators $B_N$, $A_N$ and $\mathbb{B}_N$ only.

\section{Sklyanin's  representation of Separated Variables}\label{SOV}
The eigenfunctions of the operators $B_N(u)$, $\mathbb{B}_N(u)$  and $A_N(u)$ were constructed in Refs.~\cite{Derkachov:2002tf,
Derkachov:2003qb,Belitsky:2014rba}, respectively.
In this section we present the explicit expressions for these eigenfunctions and discuss their properties.

\begin{figure}[t]
\centerline{\includegraphics[width=0.70\linewidth]{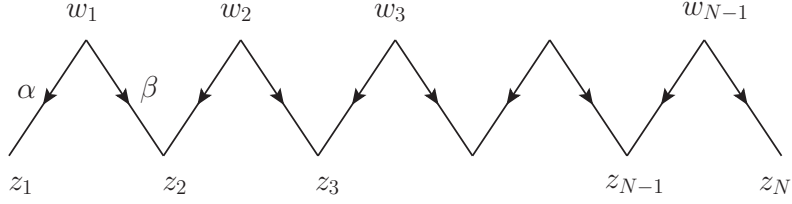}}
\label{fig:Layer}
\caption{The diagrammatic representation of the operator $\Lambda_N(u)$, $\alpha(u)=s-iu$, $\beta(u)=s+iu$. The arrow from
$\bar w$ to $z$ with the index $\alpha$ stands for the propagator $D_\alpha(z,\bar w)$, Eq.~(\ref{propagator}) }
\end{figure}
\subsection{ $B_N$-system}

\begin{figure}[t]
\centerline{\includegraphics[width=0.85\linewidth]{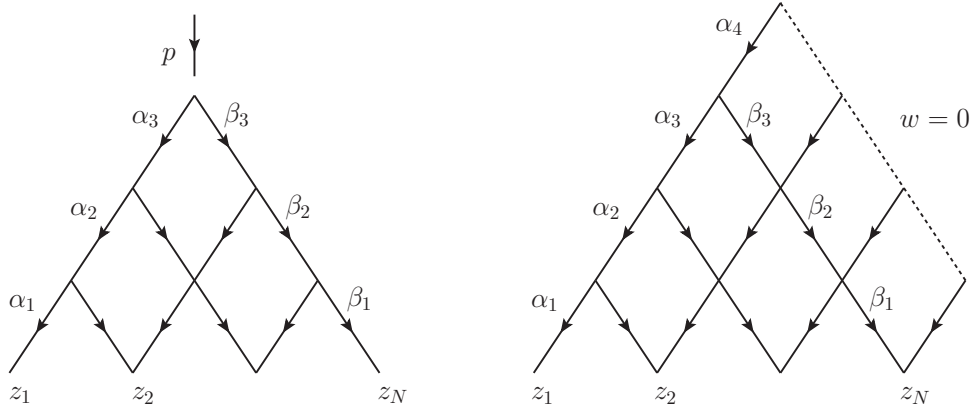}}
\caption{The diagrammatic representation of  $\Psi_B^{(N)}(p,\vec{x}|\vec{z})$,
and
 $\Psi_A^{(N)}(\vec{x}|\vec{z})$
 for $N=4$ are given on the left, respectively, right of the figure. The indices are defined as follows:
 $\alpha_k=s-ix_k$, $\beta_k=s+ix_k$. In a given layer the parallel arrows  have the same indices.
 The arrows attached to the dashed line  start from the point $w=0$.
 }
 \label{fig:AB}
\end{figure}

In order to present the result in a compact form we define  an auxiliary operator~$\Lambda_N(u)$
where $u\in \mathbb{C}$ is the spectral parameter,
This operator
maps a function of $N-1$ variables to  a function of $N$ variables according the following rule
\begin{equation}
[\Lambda_N(u)f](z_1,\ldots, z_N)=\left(\prod_{k=1}^{N-1} \int \mathcal{D} w_k\right)
\Lambda_N^{(u)}(z_1,\ldots,z_N|w_1,\ldots w_{N-1})\,
f(w_1,\ldots,w_{N-1})\,,
\end{equation}
where
\begin{align}
\Lambda_N^{(u)}(z_1,\ldots,z_N|w_1,\ldots w_{N-1})&=\prod_{j=1}^{N-1} \, D_{s-iu}(z_j,\bar w_j)\, D_{s+iu}(z_{j+1},\bar w_j)
\end{align}
and the propagator $D_\alpha$  is defined as:
\begin{align}\label{propagator}
 D_{\alpha}(z,\bar w)=\left(\frac{i}{z-\bar w}\right)^\alpha\,.
\end{align}
Note, that under conjugation the propagator  transforms as follows, $(D_\alpha(z,\bar w))^\dagger=D_{\alpha^*}(w,\bar
z)$.

The operator $\Lambda_N$ has the following properties:
%
\begin{align}\label{BLAmbda0}
&(B_N(u) \Lambda_N(u) f)(z_1,\ldots,z_N)=0\,,
\\
\label{Lambda-exchange}
&\Lambda_N(u_1)\Lambda_{N-1}(u_2)=\Lambda_N(u_2)\Lambda_{N-1}(u_1)\,.
\end{align}
%
%
The eigenfunctions of the operator $B_N$ are obtained by a consecutive application of the operators $\Lambda_k$ to
the exponential function
\begin{align}\label{B-N}
\Psi_B^{(N)}({p,\vec{x}}|\vec{z})=b_N(p)\Lambda_N(x_1)\Lambda_{N-1}(x_2) \ldots \Lambda_2(x_{N-1})\, e^{ip w}\,,
\end{align}
where $p\geq 0$ and the $x_i$ are all real~\cite{Derkachov:2003qb}.
It is convenient to fix the normalization factor $b_N(p)$ as
\begin{align}
b_N(p)=p^{Ns-\frac12} (\Gamma(2s))^{-N^2/2}.
\end{align}
It follows from (\ref{BLAmbda0}) and  (\ref{Lambda-exchange}) that
%
\begin{align}
B_N(u)\Psi_B^{(N)}({p,\vec{x}}|\vec{z})=p(u-x_1)\cdots(u-x_{N-1})\,\Psi_B^{(N)}({p,\vec{x}}|\vec{z})\,.
\end{align}
The eigenfunction is symmetric under permutation of the separated variables $\{x_1,\ldots,x_{N-1}\}$.
Since  the operator $B_N(u)$ is self-adjoint for real $u$, $(B_N(u))^\dagger= B_N(u)$,
the eigenfunctions are mutually orthogonal~\cite{Derkachov:2002tf}
\begin{align}\label{s-product-B}
\big(\Psi_B^{(N)}({p',\vec{x}'}), \Psi_B^{(N)}({p,\vec{x}})\big)=(2\pi)^{N-1}\,
\,\delta(p-p')\,
\left(\sum_{w\in S_{N-1}}
\delta(x-wx')\right)\,
\frac{\prod_{j\neq k}
\Gamma(i(x_k-x_j))
}{\prod_{k=1}^{N-1}\big[\Gamma(\alpha_{x_k})\Gamma(\beta_{x_k})\big]^N}\,,
\end{align}
where $\alpha_{x}=s-ix$,
$\beta_{x}=s+ix$
and
\begin{align}
\delta(x-x')\equiv \delta(x_1-x'_{1})\ldots \delta(x_{N-1}-x'_{{N-1}})\,.
\end{align}


\subsection{$A_N$-system}
\begin{figure}[t]
\centerline{\includegraphics[width=0.40\linewidth]{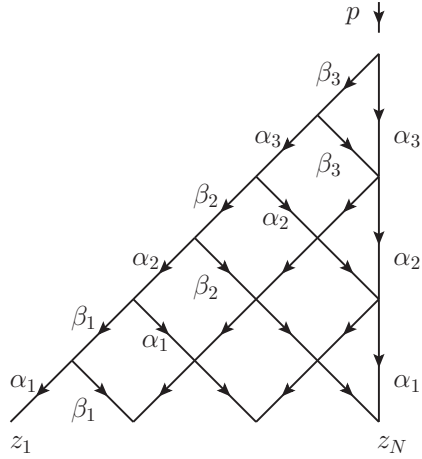}}
\caption{The diagrammatic representation of the eigenfunction $\Psi_\mathbb{B}^{(N)}(p,\vec{x}|\vec{z})$ for $N=4$.
Notation is the same as in Fig.~\ref{fig:AB}.
 }
 \label{fig:OpenB}
\end{figure}
The eigenfunctions of the operator $A_N(u)$ is constructed in a similar manner~\cite{Belitsky:2014rba}:
\begin{equation}
\Psi_A^{(N)}({\vec{x}}|\vec{z})=a_N \, \widehat \Lambda_N(x_1)\widehat \Lambda_{N-1}(x_2) \cdots \widehat \Lambda_1(x_{N})\,,
\end{equation}
where
\begin{equation}\label{aA}
a_N=
(\Gamma(2s))^{-N^2/2}%
\end{equation}
and the operator $\widehat \Lambda_N$ is defined as follows
\begin{equation}
[\widehat \Lambda_N(x)f](z_1,\ldots,z_N)=
D_{s-ix}(z_N,0)\,
[\Lambda_N(x)f](z_1,\ldots,z_N)\,.
\end{equation}
The operator satisfies equations similar to Eqs.~(\ref{BLAmbda0}) and (\ref{Lambda-exchange}):
\begin{align}
(A_N(u) \widehat\Lambda_N(u)f)(z_1,\ldots,z_N)=0\,, && \widehat\Lambda_N(u_1)\widehat\Lambda_{N-1}(u_2)=\widehat\Lambda_N(u_2)\widehat\Lambda_{N-1}(u_1)\,,
\end{align}
which ensure that
\begin{align}
A_N(u)  \Psi_A^{(N)}({\vec{x}}|\vec{z}) = (u-x_1)\cdots (u-x_N)  \Psi_A^{(N)}({\vec{x}}|\vec{z}).
\end{align}
The function $\Psi_A^{(N)}({\vec{x}}|\vec{z})$ is symmetric under permutation of the spectral parameters $x_k$, and
for the scalar product one gets~\cite{Belitsky:2014rba}
\begin{equation}\label{s-product-A}
\big(\Psi_A^{(N)}({\vec{x}'}), \Psi_A^{(N)}({\vec{x}})\big)=
(2\pi)^N\left(\sum_{w\in S_N}\delta \left(x - wx'\right)\right)
\frac{\prod_{j\neq k}
\Gamma(i(x_k-x_j))
}{\prod_{k=1}^{N}\big[\Gamma(\alpha_{x_k})\Gamma(\beta_{x_k})\big]^N}\,.
\end{equation}
%

\subsection{$\mathbb{B}_N$ system}
Since $\mathbb{B}_N(-i/2)=0$  it is convenient to redefine
the operator to get rid of this zero~\cite{Derkachov:2003qb},
\begin{equation}
\widehat{\mathbb{B}}_N(u)=\mathbb{B}_N(u)/(2u+i).
\end{equation}
This new operator is a polynomial of degree $2N-2$ in the spectral parameter $u$
and satisfies the symmetry $\widehat{\mathbb{B}}_N(u)=\widehat{\mathbb{B}}_N(-u)$.

The eigenfunctions  have the form~(\ref{B-N}) with a different type of the ``layer'' operator
$\Lambda_N\mapsto\widetilde{\Lambda}_N$
\begin{equation}\label{OB-N}
\Psi_{\mathbb{B}}^{(N)}({p,\vec{x}}|\vec{z})=c_N(p)\widetilde\Lambda_N(x_1)\widetilde\Lambda_{N-1}(x_2)
\cdots \widetilde \Lambda_2(x_{N-1})\, e^{ip w}\,,
\end{equation}
where
\begin{align}\label{cBop}
c_N(p)=p^{Ns-\frac12} \, (\Gamma(2s))^{-N(N-1/2)}\,.
\end{align}
The operator $\widetilde\Lambda_N(x)$ is defined as follows~\cite{Derkachov:2003qb}
\begin{equation}
[\widetilde\Lambda_N(x) f](z_1,\ldots,z_N)=\left( \prod_{k=1}^{N-1} \int \mathcal{D} w_{k}\right)\,
 \widetilde\Lambda_N^{(x)}(z_1,\ldots,z_N|w_1,\ldots w_{N-1})\,
f(w_1,\ldots,w_{N-1})\,,
\end{equation}
where the kernel $\widetilde\Lambda_N^{(x)}$  has the form
\begin{align}
\widetilde\Lambda_N^{(x)}(z_1,\ldots,z_N|w_1,\ldots w_{N-1}) &=D_{s-ix}(z_{N},\bar w_{N-1})
\left( \prod_{k=1}^{N-1} \int \mathcal{D} \xi_{k}\right)
\prod_{i=1}^{N-1} D_{s-ix}(z_i,\bar \xi_i)D_{s+ix}(z_{i+1},\bar \xi_i)
\notag\\
&\quad \times
\left(\prod_{k=1}^{N-2} D_{s+ix}(\xi_k,\bar w_k)D_{s-ix}(\xi_{k+1},\bar w_k)\right) D_{s+ix}(\xi_{N-2},\bar w_{N-1}).
\end{align}
This operator satisfies   equations
\begin{align}\label{open-Lambda}
(\widehat{\mathbb{B}}_N(x)\widetilde\Lambda_N^{(x)})f(z_1,\ldots,z_N)=0\,, &&
\widetilde\Lambda_N(x_1)\widetilde\Lambda_{N-1}(x_2)=\widetilde\Lambda_N(x_2)\widetilde\Lambda_{N-1}(x_1)\,, &&
\widetilde\Lambda_N(x)=\widetilde\Lambda_N(-x)\,.
\end{align}
It follows from~\eqref{open-Lambda} that
\begin{align}
\widehat{\mathbb{B}}_N(u)\Psi_{\mathbb{B}}^{(N)}({p,\vec{x}}|\vec{z}) & =
p(u^2-x_1^2)\cdots (u_2-x_{N-1}^2) \Psi_{\mathbb{B}}^{(N)}({p,\vec{x}}|\vec{z})\,
\end{align}
and $\Psi_{\mathbb{B}}^{(N)}({p,\vec{x}}|\vec{z})$ is an even function of the separated variables, $x_k$.
The scalar product of two eigenfunctions takes the form~\cite{Derkachov:2003qb}
\begin{align}\label{s-product-open}
\Big(\Psi_{\mathbb{B}}^{(N)}({p',\vec{x}'}),\Psi_{\mathbb{B}}^{(N)}({p,\vec{x}})\Big) &=(2\pi)^{N-1}\,
\,\delta(p-p')\,
\left(\sum_{{w\in S_N}}\delta(x'-wx)\right)\,
\notag\\
&\quad \times
\prod_{n=1}^{N-1}\Gamma(2ix_n) \Gamma(-2ix_n)\,
\frac{\prod_{j< k}
\Gamma(i(x_k\pm x_j))\Gamma(-i(x_k\pm x_j)) }{\prod_{k=1}^{N-1}\left[\Gamma(\alpha_{x_k})\Gamma(\beta_{ix_k})\right]^{2N}}\,,
\end{align}
where $x_k, x'_k \geq 0$, $k=1,\ldots,N-1$.

It appears quite  helpful to use a diagrammatic  representation for all objects
considered above. They can be represented in the form of  Feynman diagrams. The examples are shown in Figs.~\ref{fig:AB}
and~\ref{fig:OpenB}. In these figures the line with an arrow and  the index $\alpha$ stands for the propagator, $D_{\alpha}$,
Eq.~(\ref{propagator}), and the integration over all vertices with the measure~(\ref{measure}) is implied. Identities
like~(\ref{Lambda-exchange}) are equivalent to the equality of the corresponding diagrams and can be proved with the help of
few diagrammatical rules.

The operators $B_N(u),A_N(u), \mathbb{B}_N(u)$ are 
hermitian
operators for real $u$. Provided that they can be
extended to self-adjoint operators their eigenfunctions will form a complete system in the Hilbert space.
The direct
proof of completeness is also possible and will be given elsewhere. In particular, the completeness of the $B_N$ and
$A_N$ systems is equivalent to the completeness of the SoV representation for the Toda spin chain which was proved
by K.~Kozlowski \cite{Kozlowski:2014jka}. In what follows we take for granted that each of these systems provide a
basis in the Hilbert space $\mathcal{H}_N$.

Finally, we need two more identities for the eigenfunctions. Namely,
\begin{subequations}\label{shifts}
\begin{align}
\label{shifts-A}
A_N(x_1) \Psi_B^{(N)}({p,x_1,\ldots,x_{N-1}}|\vec{z})&  = (u+is)^N   \Psi_B^{(N)}({p,x_1+i,\ldots,x_{N-1}}|\vec{z})\,,
\\
\label{shifts-B}
B_N(x_1) \Psi_A^{(N)}({x_1,\ldots,x_{N}}|\vec{z}) & = -i (u+is)^N \Psi_A^{(N)}({x_1+i,\ldots,x_{N}}|\vec{z})\,,
\end{align}
\end{subequations}
i.e. the operators $A_N$ and $B_N$  act as shift operators on the separated variables. Since the eigenfunctions are
symmetric in $x_k$, a similar equations hold also for all others $x_k$.  The equations~(\ref{shifts}) can be derived
from the fundamental commutation relations (FCR) for the operators $A_N, B_N$, see e.g.
Refs.~\cite{Faddeev:1996iy,Sklyanin:1995bm}, or by the "gauge rotation" trick for  Lax operators,
~\cite{Derkachov:1999pz,Derkachov:2002tf}.
\section{Matrix elements}\label{sect:matrix-elements}

In this section we discuss the calculation of the scalar product of the eigenfunctions $\Psi_B^{(N)}({p,\vec{u}})$ and
$\Psi_A^{(N)}({\vec{x}})$ and the matrix element of  the shift operator,
$T_\gamma=\exp\{-\gamma S_-\}$,
 where $S_-=\sum_{k=1}^N S_-^{(k)}$ is   shift generator
\begin{equation}
T_\gamma f(z_1,\ldots,z_N)=f(z_1+\gamma,\ldots,z_N+\gamma)\,, \hskip 1cm \gamma\in \mathbb{R}\,.
\end{equation}
We introduce the following notation
\begin{align}\label{S-T}
  S^{BA}_N(p,\vec{u}; \vec{x})=\Big(\Psi_B^{(N)}({p,\vec{u}}),\Psi_A^{(N)}({\vec{x}})\Big)\,,
  &&
  T_\gamma(\vec{x},\vec{x}')=\Big(\Psi_A^{(N)}({\vec{x}'}),T_\gamma
  \Psi_A^{(N)}({\vec{x}})\Big)\,.
\end{align}
The matrix elements~(\ref{S-T}) have been calculated in~Ref.~\cite{Belitsky:2014rba}. It was achieved by going over to
the Feynman diagram representation for the quantities in question
  and subsequent evaluation of these diagrams.
 The calculation is  straightforward and will not be repeated here.
Other examples of the diagrammatic technique can be found
in~\cite{Derkachov:2002tf,Derkachov:2003qb,Derkachov:2005xn} and in Section~\ref{sect:mixed} of
the present work~\footnote{For an application of this technique
to the Toda spin chain see Ref.~\cite{Silantyev07}.}. Here we present some arguments explaining why these matrix elements
can be calculated in the closed form.

It can be shown that both these matrix elements satisfy difference equations.
 In order to derive a difference equation for the scalar product $S^{BA}_N(p,\vec{u}; \vec{x})$
we consider the matrix element of the operator $A_N(u_1)$ between the eigenstates
$\Psi_B^{(N)}({p,\vec{u}})$ and  $\Psi_A^{(N)}({\vec{x}})$. Since the function $\Psi_A^{(N)}({\vec{x}}|\vec{z})$
is the eigenfunction of the operator $A_N(u_1)$
and $A_N(u_1)$ acts as shift operator on $\Psi_B^{(N)}({p,\vec{u}})$, we get
\begin{align}
\Big(\Psi_B^{(N)}({p,\vec{u}}), A_N(u_1)\Psi_A^{(N)}({\vec{x}})\Big)
    =\prod_{k=1}^N(u_1-x_k) \Big(\Psi_B^{(N)}({p,\vec{u}}), \Psi_A^{(N)}({\vec{x}})\Big)
=\prod_{k=1}^N(u_1-x_k)S^{BA}_N(p,\vec{u}; \vec{x})
\end{align}
and
\begin{multline}
\Big(\Psi_B^{(N)}({p,\vec{u}}),  A_N(u_1)\Psi_A^{(N)}({\vec{x}})\Big)
                       = \Big(A_N(u_1)\,\Psi_B^{(N)}({p,\vec{u}}),\Psi_A^{(N)}({\vec{x}})\Big)
                      \\
=(u+is)^N \Big(\Psi_B^{(N)}({p,\vec{u}+i\vec{e}_1}),\Psi_A^{(N)}({\vec{x}})\Big)
=
(u+is)^NS^{BA}_N(p,\vec{u}+i\vec{e}_1; \vec{x}),
\end{multline}
where $\vec{u}+i\vec{e}_1=\{u_1+i,u_2,\ldots,u_{N-1}\}$. Thus we get a recurrence relation for the function $S^{BA}_N(p,\vec{u}+i\vec{e}_1;
\vec{x})$ in the variable $u_1$ of the form
\begin{equation}
(u+is)^NS^{BA}_N(p,\vec{u}+i\vec{e}_1; \vec{x})= \prod_{k=1}^N(u_1-x_k)S^{BA}_N(p,\vec{u}; \vec{x})\,.
\end{equation}
The solution of the difference equation (up to multiplication  by a periodic function of $u_1$)
has the form $\prod_{k=1}^N\Gamma(i(x_k-u_1))/\Gamma^N(s-iu_1)$. Next, proceeding  in the same way and considering
the matrix element of the operator $B_N(x_1)$ one can fix the $x_1$-dependence of $S^{BA}_N(p,\vec{u}; \vec{x})$.
 Taking into account that $S^{BA}_N(p,\vec{u}; \vec{x})$ is symmetric in
$\{x\}$ and $\{u\}$ one gets
\begin{align}\label{SAB-result}
S^{BA}_N(p,\vec{u}; \vec{x})=\frac1{\sqrt{p}}p^{-iX} 
\prod_{k=1}^N \frac{1}{\Gamma(s-ix_k)}\prod_{j=1}^{N-1}\frac{\Gamma(i(u_j-x_k))}{\Gamma(s-ix_k)\Gamma(s+iu_j)}
\times\varphi(\vec{x},\vec{u})\,,
\end{align}
where we put $X=\sum_{k=1}^N x_k$ and $\varphi(\vec{x},\vec{u})$ is  a periodic function in each variable. The
$p$-dependence follows from two relations
\begin{align}
iS_0\, \Psi_A^{(N)}({\vec{x}}|\vec{z})=-
X\, \Psi_A^{(N)}({\vec{x}}|\vec{z})
                &&\text{and}&&
                iS_0 \Psi_B^{(N)}({p,\vec{u}})|\vec{z})=i\Big(p\partial_p-1/2\Big) \Psi_B^{(N)}({p,\vec{u}})|\vec{z}).
\end{align}
In order to
fix the periodic function $\varphi(\vec{x},\vec{u})$ one can either analyse analytic properties of the function~$S^{BA}_N$
or calculate
it directly with the help of diagrammatic technique  considered here.
For the matrix element~(\ref{SAB-result}) it gives $\varphi(\vec{x},\vec{u})=1$.
Nevertheless the very possibility to obtain a matrix element by solving difference equations usually indicates that
the corresponding Feynman diagram can be calculated in a closed form.
Examples are considered in the
next section.

One has also to take care about  singularities in~(\ref{SAB-result}) arising when
$u_j\to x_k$. All $\Gamma$-functions in the numerator of~(\ref{SAB-result}) come from the integration of
the  propagator's chains (see. Ref.~\cite{Belitsky:2014rba}),
\begin{align}\label{two-prop-int}
\int D w D_{s+iu}(z,\bar w) D_{s-ix}(w,\bar \zeta) & =\frac{\Gamma(2s)}{\Gamma(\alpha_x)\Gamma(\beta_u)} \int_0^\infty
\frac{dp}{p}\frac{e^{ip(z-\bar \zeta)}}{p^{i(x-u)}} 
   =\frac{\Gamma(2s)\Gamma(i(u-x))}{\Gamma(\beta_u)\Gamma(\alpha_x)}
D_{i(u-x)}(z,\bar\zeta).
\end{align}
For $x=u$ the momentum integral diverges at the lower limit. To make sense of this integral for $x=u$ one can introduce the regulator,
$i(u-x)\to i(u-x)+\epsilon$. Technically, in order to not destroy the balance of indices that makes possible
calculation of diagrams in a closed form it is preferable to ascribe  a small positive  imaginary part to the
variables $x_k, u_k$, and replace $u_k\to \bar u_k=u_k^*$ in ~(\ref{SAB-result}). Thus we assume that $\text{Im}\,
x_k >0$, and  $\text{Im}\,u_k>0$ and write ~(\ref{SAB-result}) in the form
\begin{align}\label{SAB-Int}
S^{BA}_N(p,\vec{u}; \vec{x})=\frac1{\sqrt{p}}p^{-iX} 
\prod_{k=1}^N \frac{1}{\Gamma(s-ix_k)}\prod_{j=1}^{N-1}\frac{\Gamma(i(\bar u_j-x_k))}{\Gamma(s-ix_k)\Gamma(s+i\bar u_j)}\,.
\end{align}

The matrix element $T_\gamma(\vec{x},\vec{x}')$ was calculated in ~\cite{Belitsky:2014rba} with the help of the diagrammatic technique.
Here we only
briefly discuss the derivation of the recurrence relation. Making use of the commutation relation
\begin{equation}
[S_-,A_N(u)]=B_N(u)\,
\end{equation}
which is a consequence  of the FCR~\cite{Faddeev:1996iy}, and taking into account that $(A_N(x))^\dagger=A_N(\bar
x)$, one derives
\begin{multline}
\prod_{k=1}^N(x_1-\bar x'_k) T_\gamma(\vec{x},\vec{x}') =
        \Big(A_N(\bar x_1)\Psi_A^{(N)}({\vec{x}'}),T_\gamma\Psi_A^{(N)}({\vec{x}})\Big)=
\Big( \Psi_A^{(N)}({\vec{x}'}),A_N(x_1) T_\gamma\Psi_A^{(N)}({\vec{x}})\Big)
\\
=
\Big(  \Psi_A^{(N)}({\vec{x}'}), T_\gamma  \Big(A_N(x_1)+ \gamma B_N(x_1)\Big) \Psi_A^{(N)}({\vec{x}})\Big)=
-i \gamma(x_1+is)^N T_\gamma(\vec{x}+ i\vec{e}_1,\vec{x}')\,.
\end{multline}
A direct calculation results in the following  expression for $T_\gamma$
\begin{align}\label{T-answer}
T_\gamma(\vec{x},\vec{x}')=(\gamma+i0)^{i(X-\bar X')} e^{\frac{\pi}2(X-\bar X')}\prod_{k,j=1}^N
\frac{\Gamma(i(\bar x'_j-x_k))}{\Gamma(s-ix_j)\Gamma(s+i\bar x'_k)}\,,
\end{align}
which, as can be easily checked, satisfies the above recurrence relation.
Note also that  the operator $e^{-\gamma S_-}$ is a well--defined  on $\mathcal{H}_N$
provided that $\text{Im}(\gamma)\geq 0$. Indeed,
if $f\in \mathcal{H}_N$, then $\varphi=e^{-\gamma S_-} f$,
 ($\varphi(z_1,\ldots,z_N)=f(z_1+\gamma,\ldots,z_N+\gamma)$)
also belongs to $\mathcal{H}_N$.

\subsection{First Gustafson integral}
Expanding the eigenfunctions $\Psi_A^{(N)}$ over $\Psi_B^{(N)}$
 one obtains the following integral representation for the matrix element
$T_\gamma(\vec{x},\vec{x}')$,
\begin{equation}\label{TSS}
T_\gamma(\vec{x},\vec{x'})=\frac1{(N-1)!}\int_0^\infty dp\, e^{i\gamma p}\int_{-\infty}^\infty
\prod_{k=1}^{N-1} \frac{d u_k}{2\pi}\, \mu_N(\vec{u})\, S^{BA}_N(p,\vec{u},\vec{x})\, \Big(S^{BA}_N(p,\vec{u},\vec{x}')\Big)^\dagger\,,
\end{equation}
where the measure is defined as follows
\begin{equation}\label{measure-closed-B}
\mu_N(\vec{u})=\frac{\prod_{k=1}^{N-1}\big[\Gamma(s+iu_k)\Gamma(s-iu_k)\big]^N}{\prod_{j\neq k}
\Gamma(i(u_k-u_j))
}\,.
\end{equation}
Calculating the momentum integral and canceling common factors on both sides of~Eq.~(\ref{TSS}) one finds
\begin{align}\label{TBA}
\frac1{(N-1)!}\left(\prod_{n=1}^{N-1}\int_{-\infty}^{\infty} \frac{du_n}{2\pi} \right)\, \frac{\prod_{k=1}^N \prod_{j=1}^{N-1}
 \Gamma(i(\bar x'_k-u_j))\Gamma(i(u_j-x_k))}{
\prod_{k<j}\Gamma(i(u_k-u_j))\Gamma(i(u_j-u_k))}  & =\frac{
\prod_{k,j=1}^N \Gamma(i(\bar x'_k-x_j))}
{\Gamma(i\sum_{k=1}^{N-1}(\bar x'_k-x_k))}\,,
\end{align}
which is nothing but the first Gustafson integral~(\ref{Gustafson-I}).

Starting from the composition law for the shift operator, $T_{\gamma_1+\gamma_2}=T_{\gamma_1} T_{\gamma_2}$, one derives an
integral identity for the matrix elements~(\ref{T-answer}). It takes the form
\begin{align}\label{Iw}
\frac1{N!}\prod_{n=1}^N\int \frac{du_n}{(2\pi)} \,\zeta^{iU}\, \frac{\prod_{k,j=1}^N \Gamma(i(\bar x'_k-u_j))\Gamma(i(u_j-x_k))}{
\prod_{k<j}\Gamma(i(u_k-u_j))\Gamma(i(u_j-u_k))}  & =
\frac{\zeta^{iX'}}{(1+\zeta)^{i(\bar X'-X)}}
\prod_{k,j=1}^N \Gamma(i(\bar x'_k-x_j))\,.
\end{align}
Note, that~(\ref{Iw}) can be obtained from~(\ref{TBA}) by letting the  parameters, $x_{N}$ and $x'_N$ tend  to infinity.
Further, dividing both sides of~(\ref{Iw}) by $\zeta$ and  integrating over $\zeta$  from zero to infinity one  reproduces
the integral~(9.2) of~\cite{Gustafson}.
\section{Mixed scalar products}\label{sect:mixed}

\begin{figure}[t]
\centerline{\includegraphics[width=0.85\linewidth]{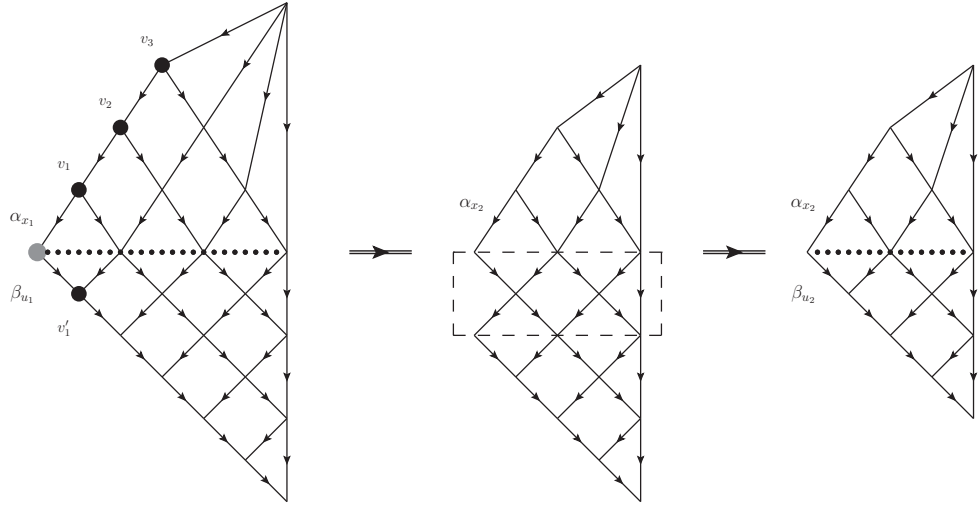}}
\caption{The diagrammatic calculation of the scalar product $\mathbb{S}^A_N(p,\vec{u}|\vec{x})$,
Eq.~(\ref{open-closed}). The leftmost diagram corresponds to the scalar product
$\mathbb{S}^A_{N=4}(p,\{u_1,u_2,u_3\}|\{x_1,x_2,x_3,x_4\})$
and the rightmost  to $\mathbb{S}^A_{N=3}(p,\{u_2,u_3\}|\{x_2,x_3,x_4\})$. The upper part of the diagrams, above the dotted line,
corresponds to the eigenfunction $A_N$  and the lower part  to the eigenfunction $\mathbb{B}_N$.
The indices are not shown explicitly. They can be easily read off the diagrams in Figs.~\ref{fig:AB} and \ref{fig:OpenB}
(notice that $\alpha_{x_1} =s-ix_1$, and $\beta_{u_1}=s+iu_1=(\alpha_{u_1})^*$).
 }
 \label{fig:Trick}
\end{figure}

In order to prove the second Gustafson integral, we consider the scalar products between the functions
of the  $\mathbb{B}$ system and the $B$ and $A$ systems.
As a first step we   derive   the recurrence relations for the matrix elements
\begin{align}\label{open-closed}
\mathbb{S}^A_N(p,\vec{u}|\vec{x})=\Big(\Psi_{\mathbb{B}}^{(N)}(p,{\vec{u}}), \Psi_A^{(N)}({\vec{x}})\Big)\,, &&
\delta(p-q)\mathbb{S}^B_N(\vec{u}|\vec{x})=\Big( \Psi_{\mathbb{B}}^{(N)}(p,{\vec{u}}), \Psi_B^{(N)}(q,{\vec{x}})\Big)\,.
\end{align}
The analysis is almost the same for both products so we consider only
$\mathbb{S}^A_N(p,\vec{u}|\vec{x})$. We start by noticing that
\begin{align}
\mathbb{B}_N(u)=B_N(-u) A_N(u)-A_N(-u) B_N(u)=(-1)^{N-1}(2u+i)\Big\{ iS_- u^{2N-2}+ O\left(u^{2N-4}\right)\Big\}
\end{align}
and
\begin{align}
\mathbb{B}_N(u)\Psi_{\mathbb{B}}^{(N)}(p,{\vec{u}}|\vec{z})=(-1)^{N-1}(2u+i) p \prod_{k=1}^{N-1}(u^2-u_k^2)
\Psi_{\mathbb{B}}^{(N)}(p,{\vec{u}}|\vec{z}).
\end{align}
Then, considering the matrix element $\Big( \Psi_{\mathbb{B}}^{(N)}(p,{\vec{u}}),
\mathbb{B}_N(x_1)\,\Psi_A^{(N)}({\vec{x}})\Big)$ and taking into account that the operator $A_N(x_1)$ annihilates
the eigenfunction $\Psi_A^{(N)}({\vec{x}})$ while $B_N(x_1)$ shifts the separated variables as given by
Eq.~(\ref{shifts-B}) one derives 
\begin{equation}\label{reccurence-B}
p\prod_{k=1}^{N-1}(x_1^2-u_k^2)\,\mathbb{S}^A_N(p,\vec{u}|\vec{x})=
i(x_1+is)^N\,\prod_{k=2}^{N}(x_1+x_k)\,\mathbb{S}^A_N(p,\vec{u}|\vec{x}+i\vec{e}_1)\,.
\end{equation}
Solving this recurrence relation and taking into account that the function $\mathbb{S}^A_N(p,\vec{u}|\vec{x})$ is
a symmetric function of the separated variables $x_1,\ldots, x_N$, we obtain
\begin{align}\label{AB-X}
\mathbb{S}^A_N(p,\vec{u}|\vec{x})= \frac1{\sqrt{p}}p^{-iX}\frac{\prod_{k=1}^N\prod_{j=1}^{N-1} \Gamma(\pm iu_j-ix_k)}{
\prod_{k=1}^N \Gamma^N(s-ix_k) \prod_{k<j}\Gamma(-i(x_k+x_j))} \times \Phi_N(u)\,.
\end{align}
Of course, there is always the possibility to multiply this expression by a periodic function in $x$. Let us for  a
moment assume  that Eq.~(\ref{AB-X})  correctly reproduces the $x$-dependence
of the function $\mathbb{S}^A_N$. Then one can see that the $\Gamma$-functions in
the second product in the denominator do not arise
from the
integration of the propagator chains. It means that the diagram which represents the matrix element
$\mathbb{S}^A_N(p,\vec{u}|\vec{x})$, see Fig.~\ref{fig:Trick}, cannot be calculated with the help of the identities
given in the~\ref{app:Diagram} only.


\begin{figure}[t]
\centerline{\includegraphics[width=0.90\linewidth]{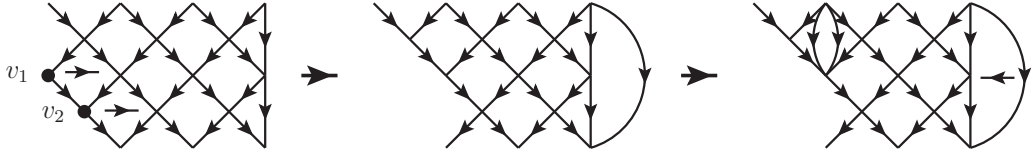}}
\caption{The diagrammatic proof of the exchange relation~(\ref{F-L-exchange}). The right diagram corresponds to the product
$\widetilde \Lambda^\dagger_{N}(v)\, \mathbb{F}_{N}(u)$. In the first step one integrates over the vertex $v_1$ and moves the
resulting line to the right with the help of the permutation relations~\ref{fig:Permutation4} and \ref{fig:Permutation3}
in \ref{app:Diagram}.
It ends up as the arched line in the middle diagram.
On the next step one repeats the same procedure for the vertex $v_2$. On the third step we insert the lines with the indices
$\pm i(u-v)$ as shown in the rightmost diagram and move them to the left and right interchanging on the  way
the indices of the two upper layers. On the final step one moves the arched line to the left. The resulting diagram corresponds
to the product $\mathbb{F}_{N-1}(u)\widetilde \Lambda^\dagger_{N}(v)$.
 }
 \label{fig:FLambda}
\end{figure}

In order to calculate~$\mathbb{S}^A_N(p,\vec{u}|\vec{x})$ we make use of the fact that  the $x$-dependence of this
function is known. Thus we have  to determine the function $\Phi_N(u)$ in Eq.~(\ref{AB-X}) only. Therefore it is
sufficient to calculate the corresponding  Feynman diagram for some appropriately chosen specialization of  the $\{x_k\}$. For
the  choice,
$x_1=u_1$, the r.h.s. in Eq.~(\ref{AB-X}) becomes singular. However, the integration over a ``free vertex'' in the diagram
for $\mathbb{S}^A_N(p,\vec{u}|\vec{x})$ (the leftmost gray vertex in  Fig.~\ref{fig:Trick} with only two propagators attached)
produces the  factor (see Eq.~(\ref{two-prop-int}))
\begin{align}
a(s+iu_1,s-ix_1)=\frac{\Gamma(2s)\Gamma(i(u_1-x_1))}{\Gamma(s+iu_1)\Gamma(s-ix_1)}\,,
\end{align}
which is also singular at $x_1\to u_1$.
Canceling the singular factor $\Gamma(i(u_1-x_1))$ on both sides one can safely put $x_1=u_1$. Since at $x_1\to u_1$  the
propagator arising due to the integration, $D_{i(u_1-x_1)}(v'_1,\bar v_1)\to 1$,
the line connecting  vertices $v_1$ and $v'_1$ disappears. The resulting diagram can be simplified as follows:
\begin{enumerate}
\item \label{i}
one integrates over the vertex
$v'_1$ and  moves the resulting line to the right with the help of the permutation relations given in \ref{app:Diagram}.
Then one  applies the same procedure to the vertices $v_1$, $v_2$ and so on.
Each integration produces  the factor $a(\alpha,\beta)$ and the successive application of the permutation relations
results in  a rearrangement of  indices.
Namely, the integration over the vertices $v'_1, v_1, v_2,\ldots$ gives the factors
$a(\alpha_{u_1},\alpha_{u_1}),\,\, a(\beta_{u_1},\alpha_{x_2}),\,\,a(\beta_{u_1},\alpha_{x_3}),\ldots $, respectively.
After these steps the upper part of the resulting diagram (the middle diagram in  Fig.~\ref{fig:Trick})
corresponds to the diagram for the eigenfunction $\Psi_A^{(N-1)}(x_2,\ldots,x_N)$.

\item The lower part of the middle diagram  in Fig.~\ref{fig:Trick} has the form
\begin{align}\label{productLambdaF}
\widetilde \Lambda^\dagger_2(u_{N-1})\cdots \widetilde \Lambda^\dagger_{N-1}(u_2)\, \mathbb{F}_{N-1}(u_1)\,,
\end{align}
where the operator $\mathbb{F}_N$ (constrained by the dashed rectangle in Fig.~\ref{fig:Trick}) has the diagrammatic
representation shown in Fig.~\ref{fig:FN}. The operators $\widetilde \Lambda^\dagger_N$ and $\mathbb{F}_N$ obey the
following exchange relation
\begin{align}\label{F-L-exchange}
 \widetilde \Lambda^\dagger_{N}(v)\, \mathbb{F}_{N}(u)=a(\alpha_u,\beta_v)a(\alpha_u,\alpha_v)\,
 \mathbb{F}_{N-1}(u)\widetilde \Lambda^\dagger_{N}(v)\,.
\end{align}
The proof of this relation is straightforward and illustrated in Fig.~\ref{fig:FLambda}. Using this relation we find
\begin{figure}[t]
\centerline{\includegraphics[width=0.55\linewidth]{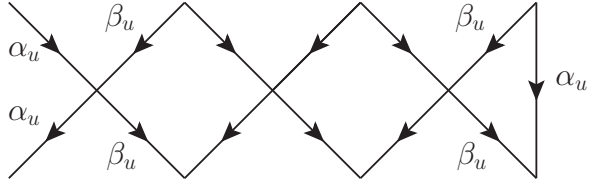}}
\caption{The diagrammatic representation of the operator $\mathbb{F}_N(u)$ (N=4)
The indices are defined as follows $\alpha_{u}=s-iu$, $\beta_{u}=s+iu$.
 }
 \label{fig:FN}
\end{figure}
%
\begin{align}
\widetilde \Lambda^\dagger_2(u_{N-1})\ldots \widetilde \Lambda^\dagger_{N-1}(u_2)\, \mathbb{F}_{N-1}(u_1)
=\prod_{k=2}^{N-1}a(\alpha_{u_1},\beta_{u_k})a(\alpha_{u_1},\alpha_{u_k})
\mathbb{F}_{1}(u_1)
\widetilde \Lambda^\dagger_2(u_{N-1})\ldots \widetilde \Lambda^\dagger_{N-1}(u_2)\,.
\end{align}
Finally, taking into account that $\mathbb{F}_1(u)(z,\bar w)=D_{s-iu}(z,\bar w)$
 and performing Fourier transform one finds  that the lower part of the diagram corresponds to the diagram for the eigenfunction
 $\Psi_{\mathbb{B}}^{(N-1)}({p,u_2,\ldots, u_{N-1}})$.
 \end{enumerate}
Thus one expresses the $N$-sites scalar product, $\mathbb{S}^A_{N}$, for  the special choice of  parameters,
$x_1=u_1$,
via the $N-1$ sites scalar product.
Collecting all  factors and taking into account the normalization coefficients~(\ref{aA}),~(\ref{cBop}) we  obtain:
\begin{align}
\frac{\mathbb{S}^A_N(p,\vec{u}|\vec{x})}{\Gamma(i(u_1-x_i))}\Big|_{u_1=x_1}
&=\mathbb{S}^A_{N-1}(p,\{u_2,\ldots,u_{N-1}\}|\{x_2,\ldots,u_{N}\})
\frac{\Gamma(2s)a(\alpha_{u_1},\alpha_{u_1})}{\Gamma(\alpha_{u_1})\Gamma(\beta_{u_1})} \prod_{k=2}^{N}a(\beta_{u_1},\alpha_{x_k})
\notag\\
&\quad \times \prod_{k=2}^{N-1}a(\alpha_{u_1},\beta_{u_k})a(\alpha_{u_1},\alpha_{u_k})\times p^{-iu_1}
\frac{\Gamma(2s)}{\Gamma(\alpha_{u_1})}\ (\Gamma(2s))^{-3N+2}\,.
\end{align}
%
Substituting $\mathbb{S}^A_{N}$ and  $\mathbb{S}^A_{N-1}$ in the form~(\ref{AB-X}) we get for
$\Phi_N(u)$
\begin{align}
\Phi_N(u_1,\ldots,u_{N-1})=\frac{1}{\Gamma^N(\alpha_{u_1})\Gamma^N(\beta_{u_1})}
\prod_{k=2}^{N-1}
\frac{1}{\Gamma(\alpha_{u_k})\Gamma(\beta_{u_k})}\Phi_{N-1}(u_2,\ldots,u_{N-1})\,.
\end{align}
Taking into account the initial condition $\Phi_1=1$ yields
\begin{align}
\Phi_N(u_1,\ldots,u_{N-1})=
\left(\prod_{k=1}^{N-1}
{\Gamma(\alpha_{u_k})\Gamma(\beta_{u_k})}\right)^{-N}\,
\end{align}
and, hence
\begin{align}\label{SAop}
\mathbb{S}^A_N(p,\vec{u}|\vec{x})= \frac1{\sqrt{p}}p^{-iX}\frac{\prod_{k=1}^N\prod_{j=1}^{N-1} \Gamma(\pm iu_j-ix_k)}{
\left(\prod_{k=1}^{N-1} \Gamma(\alpha_{u_k})\Gamma(\beta_{u_k})\prod_{k=1}^N \Gamma(\alpha_{x_k})\right)^N
\prod_{1\leq k<j<N}\Gamma(-i(x_k+x_j))} \,.
\end{align}
The calculation of the scalar product $\mathbb{S}^B_N(\vec{u}|\vec{x})$ follows along the exact same lines.
Consequently  we only state the final result
\begin{align}\label{SBop}
\mathbb{S}^B_N(\vec{u}|\vec{x})=\frac{\prod_{k=1}^{N-1}\prod_{j=1}^{N-1} \Gamma(\pm iu_j-ix_k)}{
\left(\prod_{k=1}^{N-1} \Gamma(\alpha_{u_k})\Gamma(\beta_{u_k}) \Gamma(\alpha_{x_k})\right)^N \prod_{1\leq k<j<N}\Gamma(-i(x_k+x_j))} \,.
\end{align}

In our derivation~Eq.~(\ref{SAop}) we did not consider  the possibility of multiplying the  solution~(\ref{AB-X}) of
the recursion~(\ref{reccurence-B})by a periodic function of $x$.
In order to see that Eq.~(\ref{SAop}) gives the right answer one can proceed a bit differently, finally
arriving to the same result. Namely, it can be shown by a straightforward application of the integration rules
to the diagram for  the scalar product $\mathbb{S}^A_N(p,\vec{u}|\vec{x})$, (the leftmost diagram in
Fig.~\ref{fig:Trick})
 that it can be represented
in the form $\mathbb{F}(u_1,\vec{x})\times S(u_2,\ldots, u_{N-1}|\vec{x})$, where $\mathbb{F}(u_1,\vec{x})$ is given
by a product of $\Gamma$-functions.  Since the function $\mathbb{S}^A_N(p,\vec{u}|\vec{x})$
is a symmetric function of the $\vec{u}$ variables one concludes that
$\mathbb{S}^A_N(p,\vec{u}|\vec{x})=\prod_{k=1}^{N-1}\mathbb{F}(u_k,\vec{x})\times \Psi_N(\vec{x})$. Finally,  in order to
determine $\Psi_{N}(\vec{x})$ one applies the same $``u_1\to x_1"$ trick  described above.

\subsection{Second Gustafson's integral}

The second Gustafson integral is also related to the matrix element  $T_\gamma(\vec{x},\vec{x}')$,
Eq.~(\ref{S-T}). Only this time we use an expansion in terms of the eigenfunctions $\Psi_{\mathbb{B}}^{(N)}({p,\vec{u}})$.
One obtains
\begin{equation}\label{T-SOP}
T_\gamma(\vec{x},\vec{x'})=\frac1{(N-1)!}\int_0^\infty dp\, e^{i\gamma p} \left(\prod_{k=1}^{N-1}\int_{-\infty}^\infty
 \frac{d u_k}{4\pi}\right)\,{\widetilde  \mu}_N(\vec{u})\, \mathbb{S}^{A}_N(p,\vec{u},\vec{x})\,
 \Big(\mathbb{S}^{A}_N(p,\vec{u},\vec{x}')\Big)^\dagger\,,
\end{equation}
where the measure is defined as follows, see Eq.~(\ref{s-product-open}),
\begin{equation}
\widetilde \mu_N(\vec{u})=\frac{\prod_{k=1}^{N-1}\big[\Gamma(s+iu_k)\Gamma(s-iu_k)\big]^{2N}}{\prod_{1\leq j< k<N}
\Gamma(i(u_k\pm u_j))\Gamma(-i(u_k\pm u_j))
}\,.
\end{equation}
Substituting~$\mathbb{S}^{A}_N(p,\vec{u},\vec{x})$, Eq.~(\ref{SAop}), in~(\ref{T-SOP}) and integrating over $p$ one gets,
\begin{multline}\label{AbOp}
\frac1{(N-1)!}\left(\prod_{n=1}^{N-1}\int_{-\infty}^\infty \frac{du_n}{4\pi} \right)\, \frac{\prod_{k=1}^N \prod_{j=1}^{N-1}
 \Gamma(i(x'_k\pm u_j))\Gamma(-i(x_k\pm u_j))}{\prod_{k=1}^{N-1} \Gamma(2iu_k)\Gamma(-2iu_k)
\prod_{1\leq k<j<N}\Gamma(i(u_k\pm u_j))\Gamma(-i(u_k\pm u_j))}
=\\
=\Gamma^{-1}(i(X'-X))
\prod_{k,j=1}^N \Gamma(i(x'_k-x_j))\prod_{1\leq k<j\leq N} \Gamma(i(x'_k+x'_j)) \Gamma(-i(x_k+x_j))\,.
\end{multline}
This integral, after redefining  $\alpha_k=ix'_k$, $\alpha_{N+k}=-ix_k$ and $N-1\! \to\! N$,  coincides with~(\ref{Gustafson-II}).
\vskip 5mm

Writing down the scalar product $\mathbb{S}^{A}_N(p,\vec{x},\vec{x}')$ in the $\Psi_B^{(N)}({q,\vec{u}})$ basis
one obtains the following identity
\begin{align}
\mathbb{S}^{A}_N(p,\vec{x},\vec{x}')=\frac1{(N-1)!}\left(\prod_{k=1}^{N-1}\int_{-\infty}^\infty \frac{d u_k}{2\pi}\right) \mu_N(\vec{u})\,
\mathbb{S}_N^B(\vec{x},\vec{u})\,S_{N}^{BA}(p,\vec{u},\vec{x}')\,,
\end{align}
where the measure $\mu_N(\vec{u})$ is defined in (\ref{measure-closed-B}). Substituting the explicit expressions for the
scalar products as given by  Eqs.~(\ref{SAB-Int}),~(\ref{SAop}),~(\ref{SBop}) one obtains after some redefinition the
integral~(\ref{Gustafson-III}). Let us note here that an integral similar to~\eqref{Gustafson-III} can be obtained from the
elliptic integral,  Theorem 5.3 in \cite{Spiridonov06},  by reduction  to the rational case. However, this integral and the
integral~\eqref{Gustafson-III} have a different dependence on the external parameters and can not be transformed one to the
other.

For $N=1$ the integral~(\ref{Gustafson-III}) is a special case of ~(\ref{Gustafson-I}). Conversly,  for $N>1$
the integral~(\ref{Gustafson-I}) follows from the integrals~(\ref{Gustafson-II}) and (\ref{Gustafson-III}).
If we let $\alpha_{2N+2}$ tend to infinity and  compare  the asymptotics on both sides we find
\begin{align}\label{Gustafson-II-A}
 \left(\prod_{n=1}^{N}\int_{-i\infty}^{i\infty} \frac{dz_n}{2\pi i} \right)\,
\frac{\prod_{k=1}^{2N+1} \prod_{j=1}^{N} \Gamma(\alpha_k \pm z_j)}{\prod_{k=1}^N\Gamma(\pm 2z_k)
\prod_{1\leq k<j\leq N}\Gamma(z_k \pm z_j)\Gamma(-z_k \pm z_j)}=
{2^NN!\prod_{1\leq k<j\leq 2N+1} \Gamma(\alpha_k+\alpha_j)}\,.
\end{align}
Then multiplying   both sides of~(\ref{Gustafson-III}) by
\begin{align}
\frac{\prod_{j=1}^{N+1}\prod_{k=1}^{N} \Gamma(\gamma_j \pm \beta_k)}{\prod_{k=1}^N\Gamma(\pm 2\beta_k)
\left(\prod_{1\leq k<j\leq N}\Gamma(\beta_k \pm \beta_j)\Gamma(-\beta_k \pm \beta_j)\right)}
\end{align}
and carrying out the integration  over $\beta_k$ with the help of (\ref{Gustafson-II}) and (\ref{Gustafson-II-A}) one obtains
Gustafson's first integral.

\section{ $\Psi_N \times \Psi_{N-1}$ scalar products}\label{sect:NN}
In this section we calculate another set of scalar products which lead to several new beta~-~type
integrals. Namely, we consider
the scalar products of the eigenfunctions
$\Psi_{B}^{(N)}(p,\vec{x}|\vec{z})$, $\Psi_{A}^{(N)}(\vec{x}|\vec{z})$ with  
\begin{align}\label{NU}
\Psi_{B}^{(N-1)}(p,{x_1,\ldots x_{N-2}}|\vec{z}_{N-1})\times M_\nu(z_N), &&
\Psi_{A}^{(N-1)}(p,{x_1,\ldots x_{N-1}}|\vec{z}_{N-1})\times M_\nu(z_N),
\end{align}
where $M_\nu (z)$ is   (see also \ref{app:Diagram})
\begin{equation}
M_\nu (z) = (\Gamma(2s))^{-1/2}\Gamma(s+i\nu)D_{s+i\nu}(z,0) =
(\Gamma(2s))^{-1/2} \Gamma(s+i\nu) e^{i\pi/2 (s+ i\nu)} z^{-s-i\nu}.
\end{equation}

All four scalar products can be calculated by the diagrammatic technique. The calculation  proceeds along the following
lines: one starts with the diagram $G_N$ for the $N$-point scalar product of two functions and  transforms  $G_N$ into the form
$G_N=F_N\times G_{N-1}$  with the help of the identities given in~\ref{app:Diagram}. The factor $F_N$  depends on
the spectral parameters and is given by a product of
$\Gamma$ functions.  
So one immediately gets that $G_N=F_{N}F_{N-1}\ldots F_{3}\times G_2$.
In all cases, the starting point of the recursion, the diagram $G_{N=2}$, can  be easily evaluated.

We obtained the following expressions for the  scalar products:
\begin{align}\label{A-series}
\Big(\Psi_B^{(N-1)}({p,\vec{u}})\times M_\nu, \Psi_A^{(N)}({\vec{x}})\Big)   &=
\frac{ p^{-i(\nu+X)}}{\sqrt{p}}\frac{\Gamma(\alpha_\nu)}{\Gamma(\beta_\nu)}\,
\prod_{k=1}^N\frac{\Gamma(i(x_k+\nu))}{\Gamma(\alpha_{x_k})\Gamma(\beta_{x_k})}
\notag\\
&\quad\hskip 25mm \times
\prod_{k=1}^{N-2}\frac{\Gamma(\beta_{u_k})}{\Gamma(i(u_k+\nu))}\prod_{k=1}^{N}\prod_{j=1}^{N-2}
 \frac{\Gamma(i(u_j-x_k))}{\Gamma(\alpha_{x_k})\Gamma(\beta_{u_j})},
\notag\\
\Big(\Psi_A^{(N-1)}(\vec{u})\times M_\nu, \psi_A^{(N)}({\vec{x}})\Big)   &= 2\pi
\delta(\nu+X-U)\frac{\Gamma(\alpha_\nu)}{\Gamma(\beta_\nu)}\, \frac{\prod_{k=1}^{N-1}
\Gamma(\beta_{u_k})}{\prod_{k=1}^{N} \Gamma(\beta_{x_k})} \prod_{k=1}^{N}\prod_{j=1}^{N-1}
 \frac{\Gamma(i(u_j-x_k))}{\Gamma(\alpha_{x_k})\Gamma(\beta_{u_j})}
\end{align}
and
\begin{align}\label{B-series}
\Big(\Psi_B^{(N-1)}({q,\vec{u}})\times M_\nu,\Psi_B^{(N)}({p,\vec{x}})\Big) & = {\theta(p-q)}{p^{iU+\frac12}}{q^{-iX-\frac12}}
(p-q)^{i(X-U-\nu)-1}
\notag\\
&\quad \hskip 25mm \times
\prod_{k=1}^{N-1} \frac1{|\Gamma(\alpha_{x_k})|^2} \prod_{j=1}^{N-2}
\frac{\Gamma\big(i(u_j-x_k)\big)}{\Gamma(\beta_{u_j})\Gamma(\alpha_{x_k})},
  \notag  \\
\Big(\Psi_A^{(N-1)}({\vec{x}})\times M_\nu, \Psi_B^{(N)}({p,\vec{u}})\Big)   &=\frac{p^{i(-\nu+X)}}{\sqrt{p}}
\prod_{k=1}^{N-1}\frac{1}{\Gamma(\beta_{u_k})}\frac{\Gamma(i(u_k-\nu))}{\Gamma(i(x_k-\nu))}\,
                    \prod_{k,j=1}^{N-1}\frac{\Gamma(i(x_k-u_j))}{\Gamma(\beta_{x_k})\Gamma(\alpha_{u_j})}\,.
\end{align}
Here and below $X=\sum_k x_k$, $Y=\sum_k y_k$ and $U=\sum_k u_k$.
It is  tacitly assumed that the $``+0"$ prescription is used for all $\Gamma$ functions in the numerators, i.e. $\Gamma(i(u-x))\mapsto
\Gamma(i(u-x)+\epsilon)$.

The scalar product of two functions $\Psi_1(\vec{y})$, $\Psi_2(\vec{x})$ can be written, schematically, in the form
\begin{align}\label{psiexpansion}
\big(\Psi_1(\vec{y}),\Psi_2(\vec{x})\big)=\int d\mu(\vec{u}) \big(\Psi_1(\vec{y}),\Psi_3(\vec{u})\big)
\big(\Psi_3(\vec{u}),\Psi_2(\vec{x})\big)\,,
\end{align}
where $\Psi_3(\vec{u})$ is a complete system of functions and
$\mu(\vec{u})$ is the corresponding measure.
For the functions $\Psi_k$, $k=1,2,3$ from the set
$\big\{\Psi^{(N)}_{B},\Psi^{(N)}_{A}, \Psi^{(N-1)}_{B} M_\nu,  \Psi^{(N-1)}_{A} M_\nu\big\}$
all scalar products in~\eqref{psiexpansion} are known in an explicit form
and
the right-hand side of (\ref{psiexpansion}) has
the form of a multidimensional Mellin-Barnes integral. We list below some of the integrals arising in this way.
The most interesting ones are those for which the number of external parameters minus the number of integrations is
maximal. In presenting  our results we replace the integration variables, $iu_k\mapsto u_k$,
and do the same for the external
parameters.

Since the prescription for going around the poles is fixed and integrals are convergent, the corresponding integral identities
hold for complex parameters as well. We will also sometimes shift $N\to N+1, N+2$. \vskip 3mm

\begin{itemize}
\item
The first integral arises from~(\ref{psiexpansion}) for the choice:\\
 $\big\{\Psi_1,\Psi_2,\Psi_3 \big\}=
\big\{\Psi_A^{(N-1)}({\vec{y}})\times M_\nu,\psi_A^{(N)}({\vec{x}}),\Psi_B^{(N-1)}({p,\vec{u}})\times
M_\nu\big\}$. It takes the form
\begin{align}
\frac1{N!}
\int_{-i\infty}^{i\infty} \prod_{k=1}^{N} \frac{du_k}{2\pi i} \frac{\prod_{j=1}^{N}
\prod_{k=1}^{N+1} \Gamma(y_k-u_j)
        \prod_{k=1}^{N+2} \Gamma(u_j+x_k)
        }{\prod_{k=1}^{N}\Gamma(\nu+u_k)\prod_{1 \leq i\neq j\leq N}\Gamma(u_i-u_j)}
=\frac{\prod_{k=1}^{N+2} \prod_{j=1}^{N+1}\Gamma(y_{j}+x_k)}{\prod_{k=1}^{N+2} \Gamma(\nu-x_k)}\,,
\end{align}
where $\nu=Y+X=\sum_{k=1}^{N+1} y_k+\sum_{k=1}^{N+2} x_k$, $\text{Re}\, x_k>0$, $\text{Re}\, y_k>0$
and coincides  with ~[2, Eq.(3.2)].

\item Considering $\big\{\Psi_B^{(N-1)}({p,\vec{u}})\times M_\nu, \Psi_A^{(N)}({\vec{x}}),
    \Psi_A^{(N-1)}({\vec{u}})\times M_\nu\big\}$
we get
\begin{multline}
\frac{2\pi i}{N!}\int_{-i\infty}^{i\infty} \prod_{k=1}^{N}\frac{d u_k}{2\pi i}\, \delta\left(\sum_{k=1}^N i u_k\right)
\frac{\prod_{j=1}^{N}\prod_{k=1}^{N-1}\Gamma(y_k-u_j)\prod_{k=1}^{N+1}\Gamma(x_k+u_j)}{
\prod_{1 \leq i\neq j\leq N} \Gamma(u_i-u_j)}=\\
=
\frac{\prod_{k=1}^{N+1}\Gamma\left(X - x_k\right)}{\prod_{k=1}^{N-1}\Gamma\left(X+y_k\right)}
\prod_{k=1}^{N+1}\prod_{j=1}^{N-1}\Gamma(y_j+x_k)\,,
\end{multline}
where $X=\sum_{j=1}^{N+1} x_j $ and $\text{Re}\, x_k>0, \ \text{Re}\, y_j>0$. 
For $N=2$ this identity is equivalent to the Wilson - de Branges   integral~\cite{deBranges,Wilson}.

\item The triple
$\big\{ \Psi_A^{(N-1)}({\vec{y}})\times M_\nu,\, \Psi_B^{(N)}({p,\vec{x}}),\,\Psi_B^{(N-1)}({q,\vec{u}})\times M_\nu\big\}$
gives rise to
\begin{multline}
\frac{1}{N!}\int_{-i\infty}^{i\infty} \prod_{k=1}^{N}\frac{d u_k}{2\pi i}\frac{\Gamma(\nu-X-U)}{\Gamma(\nu+Y-U)}
\frac{\prod_{k=1}^{N+1}\prod_{j=1}^{N}\Gamma(y_k-u_j)\Gamma(x_k+u_j)}{
\prod_{1 \leq i\neq j\leq N} \Gamma(u_i-u_j)}
\\
=
\prod_{k=1}^{N+1}\frac{\Gamma(\nu-x_k)}{\Gamma(\nu+y_k)}
\frac{\prod_{k,j=1}^{N+1}\Gamma(y_j+x_k)}{\Gamma(X+Y)}\,,
\end{multline}
where $\text{Re}\, x_k>0$, $\text{Re}\,y_k>0$,  and $\text{Re}\,\nu >\text{Re} X$. For $N=1$ it is equivalent to
the second Barnes Lemma, while for  general $N$ it is  a modification of   Gustafson's  integral~(\ref{Gustafson-I}).

\item For  $\big\{\Psi_B^{(N)}({p,\vec{y}}),\,
    \Psi_A^{(N)}({\vec{x}}),\,\Psi_A^{(N-1)}({\vec{u}})\times M_\nu\big\}$ one finds
\begin{multline}
\frac{1}{N!}\int_{-i\infty}^{i\infty} \prod_{k=1}^{N}\frac{d u_k}{2\pi i}\frac{\Gamma(s-X-U)}{\Gamma(s+X+U)}\prod_{k=1}^N\frac{\Gamma(X+U-y_k)}{\Gamma(X+U-u_k)}
\frac{\prod_{j=1}^{N}\prod_{k=1}^{N}\Gamma(y_k-u_j)\prod_{k=1}^{N+1}\Gamma(x_k+u_j)}{
\prod_{1 \leq i\neq j\leq N} \Gamma(u_i-u_j)}
=\\
=
\prod_{k=1}^{N+1}\frac{\Gamma(s-x_k)}{\Gamma(s+x_k)}\prod_{k=1}^{N}\frac{\Gamma(s-y_k)}{\Gamma(s+y_k)}
\prod_{k=1}^{N+1}\prod_{j=1}^{N}\Gamma(y_j+x_k),
\end{multline}
where $\text{Re}\, x_k>0$, $\text{Re}\,y_k>0$, $\text{Re}\,X>\text{Re}\,y_k$ and $\text{Re}\,s>\text{Re} X$. Again, for $N=1$ it
reduces to  Barnes' second lemma.

\item The last integral arises from
    $\big\{\Psi_B^{(N)}({p,\vec{y}}),\,\Psi_A^{(N)}({\vec{x}}),\,\Psi_B^{(N-1)}({p,\vec{u}})\times M_\nu\big\}$
\begin{multline}
\frac1{(N-1)!}\int_{-i\infty}^{i\infty} \frac{d\nu}{2\pi i} \int_{-i\infty}^{i\infty} \prod_{k=1}^{N-1}\frac{d u_k}{2\pi i}
\frac{\Gamma(s-X-\nu)}{\Gamma(s+X+\nu)} \frac{\prod_{k=1}^{N+1} \Gamma(\nu+X -x_k)}{\prod_{k=1}^{N-1} \Gamma(\nu+X +u_k)}
\frac{\Gamma(Y-\nu)\Gamma(\nu+U+X-Y)}{\Gamma(X+U)}
\\
\times\frac{\prod_{j=1}^{N-1}\prod_{k=1}^{N}\Gamma(y_k-u_j)\prod_{k=1}^{N+1}\Gamma(x_k+u_j)}{
\prod_{1\leq i\neq j\leq N-1} \Gamma(u_i-u_j)}
=\prod_{k=1}^{N+1}\frac{\Gamma(s-x_k)}{\Gamma(s+x_k)}\prod_{k=1}^{N}\frac{\Gamma(s-y_k)}{\Gamma(s+y_k)}
\prod_{k=1}^{N+1}\prod_{j=1}^{N}\Gamma(y_j+x_k)
\,.
\end{multline}
Here $\text{Re}\, x_k >0$,  $\text{Re}\, y_k >0$, $\text{Re}\,(X-Y)>0$ and
$\text{Re}\,(s-X)>0$. For $N=1$ it is equivalent to  Barnes' second Lemma.
\end{itemize}

\vskip 3mm

\section{Summary}\label{sect:summary}
The eigenfunctions of the matrix elements of the monodromy matrix (for both  the closed and open spin chains) provide
convenient bases for the study of the spectral problem for the corresponding spin magnets. Remarkably these
eigenfunctions  can be constructed explicitly as multivariable integrals and  represented by  Feynman diagrams of
a certain type. The  scalar products between the different eigenfunctions can be  calculated with the help of the
diagrammatic technique and, as a rule, are given by a product of gamma  functions with arguments depending on
the parameters labeling the eigenfunctions (separated variables). In the SoV representations the scalar product or
matrix elements take the form of multidimensional Mellin--Barnes  integrals. Studying different scalar products  we
succeeded  to reproduce
 all relevant integrals  in Ref.~\cite{Gustafson,Gustafson92} except for
 [1,Eq.~(9.6)]
 and [2,Ea.~(5.4)] in Ref.~\cite{Gustafson92} and derived several new
integrals which did not follow  from those of Gustafson's.

In this work we have considered only the homogeneous spin chains. However, the eigenfunctions can be constructed in
a similar way and for   general case of  inhomogeneous spin chains with impurities~\cite{Derkachov:2002tf}.
Gustafson's integrals~(\ref{Gustafson-I}) and (\ref{Gustafson-II}) are not sensitive to all these modifications. At the
same time we expect that  inclusion of  additional parameters (spins and impurities) into consideration could lead
to modification of the integrals given in Section~\ref{sect:NN}.

Our approach can be extended to  noncompact $SL(2,\mathbb{C})$ spin magnets~\cite{Derkachov:2001yn} resulting in another
extension of Gustafson's integrals. Some insight into the possible structure of such integrals can be gained
from~\cite{Bazhanov:2013bh,Kels:2013ola,Kels:2015bda}. We also expect that considering  trigonometric and elliptic spin chains
gives rise to new $q-$beta and elliptic Gustafson type integrals.

\ack  The authors are grateful to Karol Kozlowski and Vyacheslav Spiridonov for  fruitful discussions and correspondence. We
also  express our gratitude to  Hjalmar Rosengren and the referee for bringing the  reference~\cite{Spiridonov06} to our
attention.
The authors are deeply indebted to referee for a careful reading
of the manuscript and suggestion of many amendments.
This study was supported by the Russian Science Foundation (S.~D.), project
$\text{N}^{\text{o}}$ 14-11-00598,  and by Deutsche Forschungsgemeinschaft (A.~M.),  grant MO~1801/1-1.

\appendix

\section*{Appendices}
\addcontentsline{toc}{section}{Appendices}

\renewcommand{\theequation}{\Alph{section}.\arabic{equation}}
\renewcommand{\thetable}{\Alph{table}}
\setcounter{section}{0}
\setcounter{table}{0}
\section{Diagram technique}\label{app:Diagram}

In this Appendix we present the basic elements of the diagram technique which was used throughout the paper.
The propagator
\begin{equation}
D_\alpha(z,\bar w) =\left(\frac i{z-\bar w}\right)^\alpha
   =\frac{1}{\Gamma(\alpha)}\,\int_0^{\infty} dp\,e^{ip\,(z-\bar w)}\,p^{\alpha-1} 
\end{equation}
is shown by an arrow directed from $\bar w$ to $z$ with the
index $\alpha$ attached to it. Under complex conjugation it behaves as:
$(D_\alpha(z,\bar w))^*=D_{\alpha^*}(w,\bar z)$.

There are several useful identities involving propagators:
\begin{enumerate}
\item Chain rule: the integral of two propagators is  again a propagator:
\begin{figure}[H]
\centerline{\includegraphics[width=0.55\linewidth]{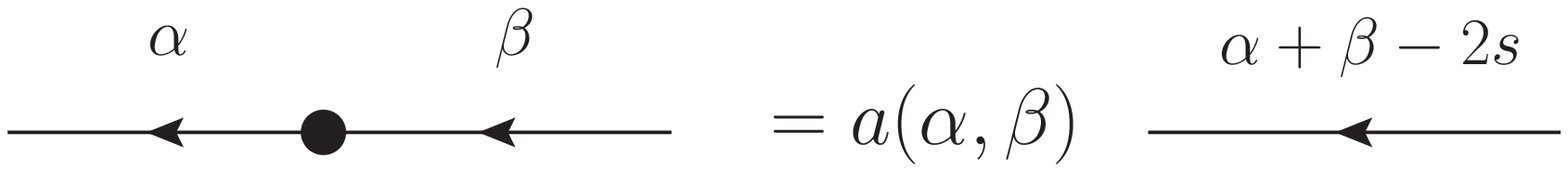}}
\end{figure}
%
\begin{align}
\int \mathcal{D} w\,D_\alpha(z,\bar w) D_\beta(w,\bar \zeta)=a(\alpha,\beta) D_{\alpha+\beta-2s}(z,\bar \zeta)\,,
\end{align}
where
\begin{align}
a(\alpha,\beta)=\frac{\Gamma(2s) \Gamma(\alpha+\beta-2s)}{\Gamma(\alpha)\Gamma(\beta)}\,.
\end{align}

\item Permutation relation:
\begin{figure}[H]
\centerline{\includegraphics[width=0.70\linewidth]{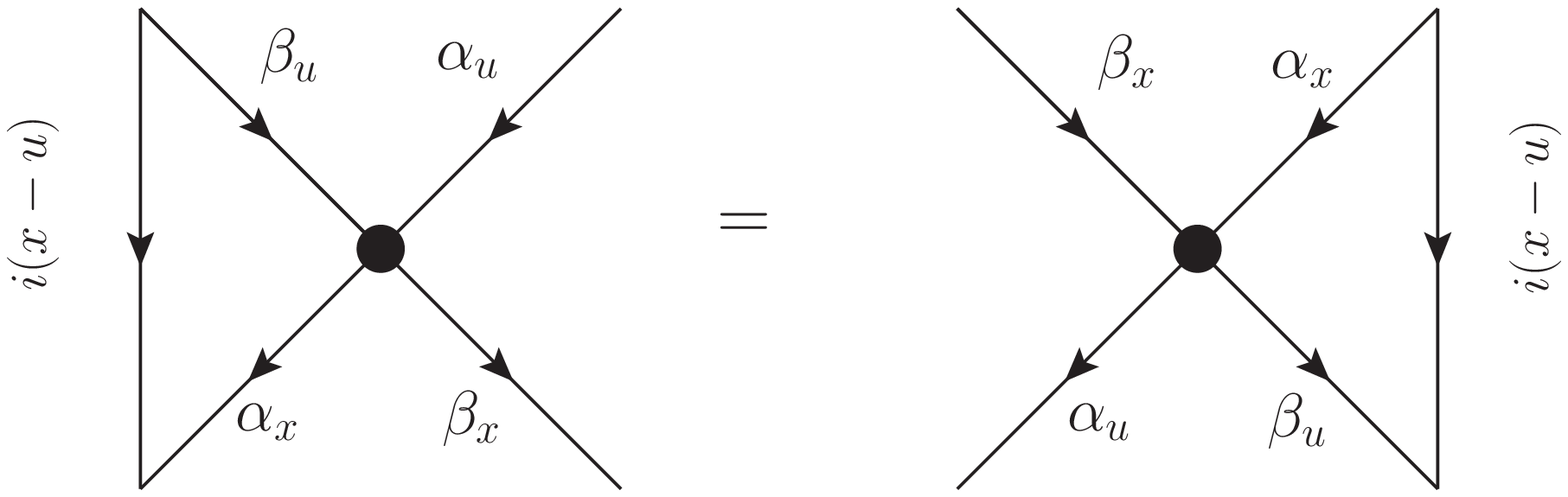}}
 \label{fig:Permutation4}
\end{figure}

\item  Reduced permutation relation:
\begin{figure}[H]
\centerline{\includegraphics[width=0.70\linewidth]{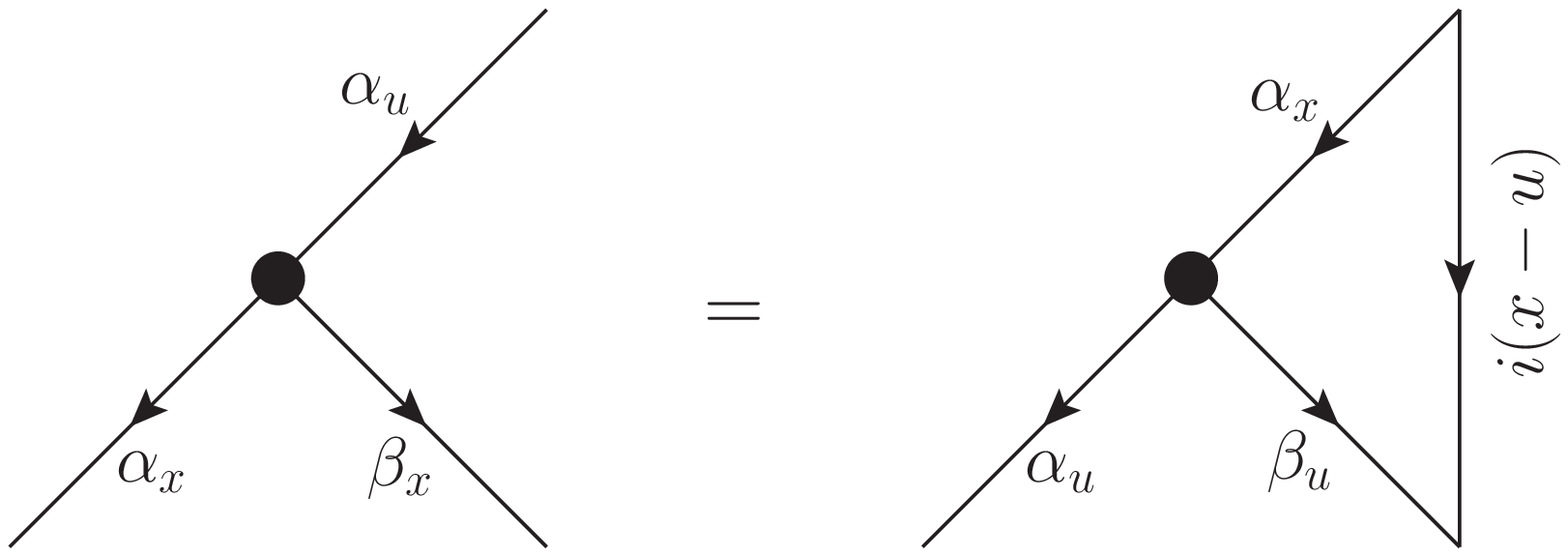}}
 \label{fig:Permutation3}
\end{figure}

\item The propagator identity:
%
\begin{figure}[H]
\begin{minipage}{1.0\textwidth}
\centerline{\includegraphics[width=0.450\linewidth]{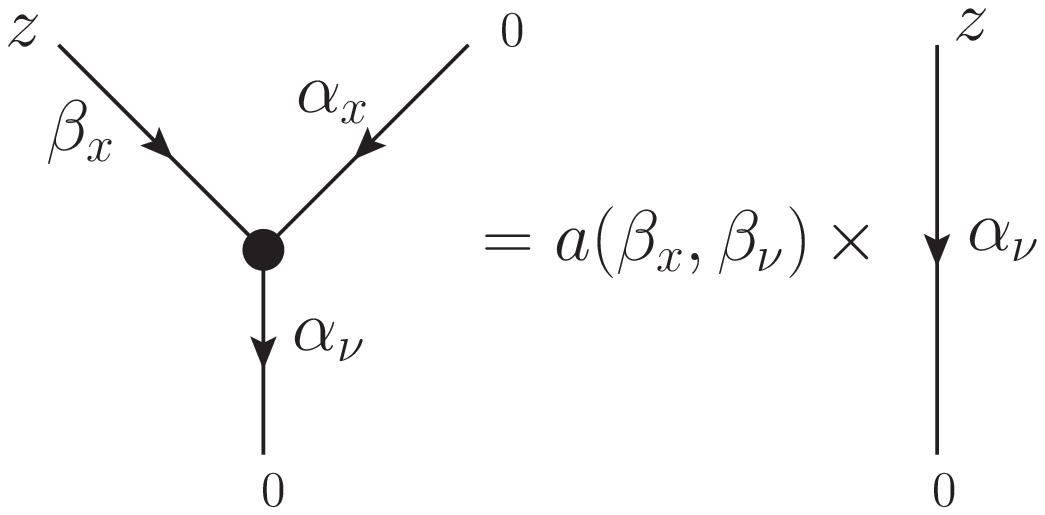}}
\end{minipage}
 \label{fig:z00}
\end{figure}
%
\end{enumerate}
All these identities can be easily checked by going over to the momentum representation. \vskip 5mm \noindent There
are two standard bases in the (one-particle) Hilbert space $\mathcal{H}$:
\begin{itemize}
\item The plane waves: $E_p(z) =p^{s-1/2} e^{ip z}/{\Gamma^{1/2}(2s)}$, $p>0$:
\begin{equation}
(E_{p'}, E_p)=\int \mathcal{D} z \, E_p(z) \overline{E_{p'}(z)}= \delta(p-p')\,.
\end{equation}
\item powers:
\begin{equation}\label{Mnu}
M_\nu (z) = (\Gamma(2s))^{-1/2}\Gamma(s+i\nu)D_{s+i\nu}(z,0) =
(\Gamma(2s))^{-1/2} \Gamma(s+i\nu) e^{i\pi/2 (s+ i\nu)} z^{-s-i\nu}
,
\end{equation}
where $\nu\in \mathbb{R}$,
\begin{equation}
(M_{\nu'}, M_\nu)=\int \mathcal{D} z \,M_\nu(z) \overline{M_{\nu'}(z)}= 2\pi \delta(\nu-\nu')\,.
\end{equation}
For the transition matrix element one obtains
\begin{equation}
\Big(M_\nu, E_p\Big)=p^{-i\nu-1/2}\,.
\end{equation}
\end{itemize}


\end{document}